\documentclass[sigconf]{acmart}

\usepackage{enumitem}
\usepackage{hyperref}
\usepackage{tabularx}
\usepackage{xurl}  
\usepackage{graphicx}
\usepackage{amsmath}
\usepackage{hyperref}
\usepackage{multirow, multicol, makecell}
\usepackage[breakable,skins]{tcolorbox}

\newcommand{\palatino}{\fontfamily{ppl}\selectfont}

\newcommand{\thickhline}{
    \noalign{\hrule height 2pt}
}

\newcommand{\TCD} {Toxic Conversations Dataset}
\newcommand{\DTD} {Derailed Toxic Dataset}
\newcommand{\NTD} {Non-Toxic Conversations Dataset}

\definecolor{colortitlebg}{HTML}{1D1D1D}
\definecolor{colorcommentbg}{HTML}{E2F2DF}
\newenvironment{promptbox}[2][]{
    \begin{tcolorbox}[title={#2},
    fonttitle={\palatino\bfseries}, 
    enhanced jigsaw, 
    colbacktitle={colortitlebg},
    arc=2pt,
    opacityframe=0,
    boxrule=0.4mm,
    opacityframe=1, 
    colback={colorcommentbg},#1,
    breakable]
}{\end{tcolorbox}}

\newcommand{\change}[1]{\textcolor{black}{#1}}

\title{Toxicity Ahead: Forecasting Conversational Derailment on GitHub}

\author{Mia Mohammad Imran}
\affiliation{%
  \institution{Missouri University of Science and Technology}
  \city{Rolla}
  \state{MO}
  \country{USA}
}
\email{imranm@mst.edu}

\author{Robert Zita}
\affiliation{%
  \institution{Elmhurst University}
  \city{Elmhurst}
  \state{IL}
  \country{USA}
}
\email{rzita8729@365.elmhurst.edu}

\author{Rahat Rizvi Rahman}
\affiliation{%
  \institution{Virginia Commonwealth University}
  \city{Richmond}
  \state{VA}
  \country{USA}
}
\email{rahmanr12@vcu.edu}

\author{Preetha Chatterjee}
\affiliation{%
  \institution{Drexel University}
  \city{Philadelphia}
  \state{PA}
  \country{USA}
}
\email{preetha.chatterjee@drexel.edu}

\author{Kostadin Damevski}
\affiliation{%
  \institution{Virginia Commonwealth University}
  \city{Richmond}
  \state{VA}
  \country{USA}
}
\email{kdamevski@vcu.edu}

\begin{document}

\begin{abstract}
Toxic interactions in Open Source Software (OSS) communities reduce contributor engagement and threaten project sustainability. Preventing such toxicity before it emerges requires a clear understanding of how harmful conversations unfold. However, most proactive moderation strategies are manual, requiring significant time and effort from community maintainers. To support more scalable approaches, we curate a dataset of 159 derailed toxic threads and 207 non-toxic threads from GitHub discussions.
Our analysis reveals that toxicity can be forecasted by tension triggers, sentiment shifts, and specific conversational patterns.

We present a novel Large Language Model (LLM)-based framework for predicting conversational derailment on GitHub using a two-step prompting pipeline. First, we generate \textit{Summaries of Conversation Dynamics} (SCDs) via Least-to-Most (LtM) prompting; then we use these summaries to estimate the \textit{likelihood of derailment}. Evaluated on Qwen and Llama models, our LtM strategy achieves F1-scores of 0.901 and 0.852, respectively, at a decision threshold of 0.3, outperforming established NLP baselines on conversation derailment. External validation on a dataset of 308 GitHub issue threads (65 toxic, 243 non-toxic) yields an F1-score up to 0.797. Our findings demonstrate the effectiveness of structured LLM prompting for early detection of conversational derailment in OSS, enabling proactive and explainable moderation.
\end{abstract}




\copyrightyear{2026}
\acmYear{2026}

\maketitle

\section{Introduction}

% toxicity is bad for GitHub project health
Toxic language undermines the health of online communities, including those centered around software projects.
A 2024 GitHub survey reported that 64.23\% of developers experienced or witnessed negative interactions~\cite{GitHub_GitHub_Open_Source_2024}, a slight increase from the 60.0\% recorded in 2017~\cite{GitHubOpenSourceSurvey2017}; notably, 21.4\% reported that such interactions led them to stop contributing.
% toxicity detection is post-hoc
Despite the increasing recognition of the negative impact of toxic interactions, to our best knowledge, all existing toxicity detection methods are post-hoc~\cite{sarker2020benchmark, sarker2023automated, mishra2024exploring, raman2020stress}, identifying toxic content only after it appears.
While post-hoc detection can mitigate some of the damage caused by toxic interactions, it fails to prevent the initial harm and may allow negative behaviors to persist unchecked for extended periods. A reactive approach not only delays intervention but also burdens community moderators and risks alienating contributors who might have otherwise remained engaged. Consequently, there is a pressing need for proactive solutions that can anticipate and preemptively address potential toxicity~\cite{li2021code}.

% Proactive moderation is better than reactive
The primary strategy of proactive moderation involves human moderators actively engaging with ongoing conversations to prevent them from devolving into toxic behavior or, at the very least, to swiftly address any negativity should it arise and before it escalates any further~\cite{schluger2022proactive, chang2019trouble}.
While effective, manual proactive moderation in OSS is impractical since moderators need to continuously monitor ongoing conversations across several communication channels (e.g., issues, chats, discussion boards). 
On the other hand, automated moderation offers scalability but demands a deep understanding of community norms and context. 
However, unlike platforms such as X (formerly Twitter) or Reddit, GitHub exhibits more subtle toxic behaviors, such as entitlement, miscommunication, or resistance to new practices, rather than overt aggression~\cite{hsieh2023nip}. Moreover, domain-specific terms in software engineering (e.g., `kill,' `dead,' and `dump') can pose challenges to generic automated toxicity detection~\cite{sarker2023automated, imran2024shedding}.

In this paper, we investigate how GitHub conversations derail into toxicity and present an automated approach for predicting such derailment. Leveraging recent advances in LLMs, we introduce a novel framework that integrates advanced prompting techniques to support proactive moderation in OSS communication channels. More specifically, we address these three research questions:

\noindent
\textit{\textbf{RQ1:} What are the characteristics and patterns of conversational derailment in GitHub discussions?}

Effective prediction requires deep understanding of how technical conversations deteriorate. We curated a dataset of 159 derailed toxic and 207 non-toxic GitHub conversations, annotated with derailment and toxicity points. Then, we empirically analyze temporal dynamics, linguistic patterns, and contextual triggers that precede toxicity. Our findings show that derailment on GitHub often precedes toxicity by a narrow margin; the median distance between the first derailment point and the first toxic comment is just 3 comments, and 64\% of toxic comments occurred within 24 hours. We identify a set of consistent early warning signals at derailment points, including elevated use of reasoning terms (e.g., "because", "since"), WH-questions (e.g., "why", "how"), and second person pronouns (e.g., "you", "your"), as well as tones like \textit{Frustration} or \textit{Impatience}.

\noindent
\textit{\textbf{RQ2:} Can Large Language Models effectively predict conversational derailment on GitHub?}

We develop and evaluate a novel LLM-based framework that generates interpretable conversation summaries to predict derailment. More specifically, we leverage \textit{Least-to-Most} prompting to generate high-level \textit{Summaries of Conversation Dynamics} (SCDs). These summaries abstract away technical details to highlight interaction patterns, emotional tone, and rhetorical shifts. Based on these SCDs, we predict the likelihood of a conversation derailing into toxicity. 
\change{To enable forecasting rather than detection, our approach for SCD generation only considers comments that occur before the first toxic remark, ensuring that predictions are based solely on pre-toxicity context. This design aligns with prior derailment forecasting formulations~\cite{chang2019trouble}.}

Our framework achieves F1-scores of 0.901 (Qwen) and 0.852 (Llama), significantly outperforming established baselines including CRAFT~\cite{chang2019trouble} and Hua et al.'s few-shot SCD~\cite{hua2024did} approaches while maintaining interpretability for human moderators.  Through ablation studies, we find that sentiment evolution and tension triggers are the most critical components for prediction accuracy.

\noindent
\textit{\textbf{RQ3:} To what extent does the proposed LLM-based derailment prediction approach generalize to independent GitHub datasets?}

We validate our approach on independent GitHub data to establish confidence in its broader applicability across different communities and time periods. External validation on the Raman et al.'s~\cite{raman2020stress} dataset (308 threads, 65 toxic and 243 non-toxic) shows reasonable generalizability, with our LtM strategy achieving F1-scores of 0.797 (Qwen) and 0.776 (Llama). The approach outperforms baselines on external data despite different collection methodologies, time periods, and class distributions, suggesting that our method captures broader patterns of GitHub conversational dynamics.

\noindent
\textbf{Paper Contributions.} Our investigation yields several key contributions. We provide the empirical characterization of GitHub conversational derailment patterns, revealing predictable deterioration signals including temporal proximity (median 3 comments before toxicity), distinctive linguistic markers, and common triggers. We introduce a novel LLM-based framework using structured Least-to-Most prompting to generate explainable Summaries of Conversation Dynamics, achieving F1-scores of 0.901 while maintaining interpretability. Finally, we provide actionable insights for OSS communities seeking to implement proactive moderation strategies.

% The remainder of this paper is organized as follows: Section 2 provides a GitHub derailment example, Section 3 describes our dataset curation, Section 4 analyzes derailment characteristics, Section 5 presents our LLM-based framework, results and ablation studies, Sections 7 provide external validation, and Sections 7-10 discuss recommendations, related work, threats to validity, and conclusions, respectively. 

Our study's datasets, scripts, and output logs are publicly available online to facilitate reproducible research~\cite{zenodo_15725618}.

\section{Example of Conversational Derailment on GitHub}\label{sec:example}

\begin{figure}[t]
\centering
\includegraphics[width=0.98\linewidth]{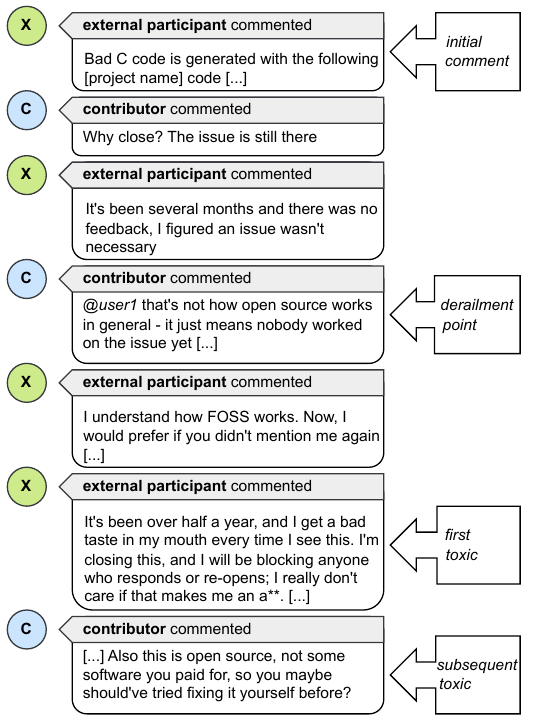}
\caption{Example of a toxic conversation on GitHub.}
\label{fig:motivating_example}
\end{figure}

In online forums, toxicity often occurs after identifiable signs in the previous comments by the discussion participants~\cite{chang2019trouble}. In this research, we focus on understanding the early signs that a conversation will turn toxic on GitHub issues and pull request discussions. These preceding comments, where it becomes clear that the conversation has moved away from being productive and taken a turn towards negativity, are called {\em derailment points}~\cite{zhang2018conversations}. 

Overall, toxic conversations often contain the following identifiable elements: 1) a conversation-initiating comment, 2) a derailment point comment, 3) a first toxic comment, and 4) (zero or more) subsequent toxic (or non-toxic) comments. Figure~\ref{fig:motivating_example} shows an example of a toxic conversation, highlighting these different structures. In this conversation between an OSS project contributor and an external participant (i.e., someone who has never made a commit to the repository), the contributor derails the conversation by making a mocking comment. The external participant responds with frustration and then makes a toxic, insulting remark. This is followed by another toxic comment, this time made by the contributor.

\section{Dataset}

Understanding the characteristics of derailed conversations is crucial for developing effective intervention strategies. To facilitate that, we curate two datasets comprising conversations from GitHub issues and pull requests: one containing derailed toxic threads and the other consisting of non-toxic threads. We describe our annotation process 
in detail, with an emphasis on ensuring label reliability and curating examples that are both representative of OSS discourse and suitable for downstream analysis.
%for each dataset, where our primary goal was to include high-quality annotated and representative examples of GitHub conversations.

\begin{table}
\caption{Definitions and examples of uncivil tone-bearing discussion features (TBDF).}
\centering
%\footnotesize
\small
\begin{tabular}{|>{\raggedright\arraybackslash}m{1.2cm}|>{\raggedright\arraybackslash}m{3cm}|>{\raggedright\arraybackslash}m{3cm}|}
\hline
\textbf{TBDF} & \textbf{Definition} & \textbf{Example}\\ \thickhline

Bitter Frustration
& Expressing strong frustration, displeasure, or annoyance	
& \textit{No answer, no reaction, what kind of support is that.}\\ \hline

Impatience
& Expressing dissatisfaction due to delays	
& \textit{Issue not fixed in 30 days? Must be gone!}\\ \hline

Mocking
& Ridiculing or making fun of someone in a disrespectful way 
& \textit{Legend says this issue will still exist even on the end of mankind.}\\ \hline

Irony
& Using language to imply a meaning that is opposite to the literal meaning, often sarcastically 
& \textit{Maybe you should actually write that down somewhere. You know, like in the documentation.} \\ \hline

Vulgarity
& Using offensive or inappropriate language 
& \textit{Who cares, same sh*t.} \\ \hline

Threat
& Issuing a warning that implies a negative consequence
& \textit{Any further responses will result in you being blocked from the repo entirely.} \\ \hline

Entitlement
& Expecting special treatment or privileges 
& \textit{that's how good we are. I don't want your contribution. [...]} \\ \hline

Insulting
& Making derogatory remarks towards another person or project
& \textit{This looks like it was done by a 5 year old.} \\ \hline

Identity attacks/\newline Name-calling
& Making derogatory comments based on race, religion, gender, sexual orientation, or nationality
& \textit{I would not be surprised if this database is maintained by the [nationality].} \\ \hline
\end{tabular}%
\label{tab:uncivil_features}
\end{table}

\subsection{{\TCD}} 
\label{toxic-dataset}
We start with a dataset recently released by Ehsani et al.~\cite{ehsani2024incivility}, which focuses on incivility in GitHub conversations. Toxicity is a subset of incivility, focusing on harmful language, while incivility more broadly includes behaviors that undermine constructive discussion~\cite{ferreira2021shut, sadeque2019incivility}. More specifically, incivility is defined as \textit{``features of discussion that convey an unnecessarily disrespectful tone toward the discussion forum, its participants, or its topics"}~\cite{coe2014online}, while toxicity is defined as \textit{``rude, disrespectful, or unreasonable language that is likely to make someone leave a discussion"}~\cite{perspectiveapi}.  Therefore, leveraging this incivility dataset provides an appropriate starting point for identifying toxic interactions in GitHub threads.

We use an LLM-aided model-in-the-loop annotation approach to identify the uncivil comments that are also toxic~\cite{bartolo2020beat}. 
Recent research shows that such a model-in-the-loop annotation methodology works well for this type of data, including hate and violent speech detection tasks~\cite{sanyal2024machines, zendel2024enhancing, gilardi2023chatgpt, nguyen2024human, wang2024human, zhu2024apt, jahan2024finding}. This methodology has been leveraged in software engineering toxicity detection as well~\cite{imran2025silent, mishra2024exploring}. In our annotation, we leveraged the prompt released by Imran et al. for toxicity detection in GitHub bug reports~\cite{imran2025silent}.
For each uncivil comment identified in Ehsani et al.’s dataset, we used GPT-4o to assess whether the comment was toxic, providing the full preceding conversation context up to that point. GPT-4o labeled 832 comments across 273 threads as toxic. To validate these predictions, two authors independently reviewed each flagged comment. The initial inter-annotator agreement was 0.78 (Cohen’s Kappa). The annotators resolved all disagreements through in-person discussion in order to finalize the toxicity annotations.

Ehsani et al.'s dataset is based on 404 locked conversations (issues and PRs) on GitHub where the reason they were locked is listed as `too heated,' `spam,' or `off-topic.' These 404 conversation threads contain 5961 comments annotated with various categories of uncivil TBDFs (Tone Bearing Discussion Features). The definitions and examples of the incivility-related TBDFs are shown in Table~\ref{tab:uncivil_features}. Since the focus of our study is to identify conversations that derail into toxicity, we excluded conversations in which toxicity occurred in the initial post (i.e., we cannot predict derailment without processing the initial post).
% or where no clear derailment point could be identified, i.e., cases lacking a meaningful progression toward toxicity. 
This step resulted in a dataset consisting of 175 toxic GitHub threads.

%Following this step, we further excluded conversations where the actual post in the conversation was marked as toxic (i.e., we cannot predict derailment without processing the initial post). 

\subsection{\NTD} 
\label{nontoxic-dataset}
To compare toxic conversations with ordinary, non-toxic GitHub issues and pull requests, we collected a sample of non-toxic dataset using the toxic dataset as a reference point. Specifically, we gathered 15 threads that were posted immediately before and 15 threads immediately after each toxic thread within the same repository, maintaining temporal and project-local continuity. We applied several exclusion criteria, removing: 1) non-English conversations, 2) threads with no comments after the initial post, and 3) locked threads marked as "resolved" (i.e., locked for reasons unrelated to toxic discourse). 

% From those collected posts, two authors manually examined each of them to ensure it is not toxic, using the definition in section \ref{toxic-dataset}. 
From the collected posts, two annotators (authors of this paper) randomly selected posts one by one and verified their non-toxic nature, using the definition in section \ref{toxic-dataset}. 
% Two authors conducted this annotation using the above-mentioned toxicity definition. 
The annotators continued this process until the number of \textit{non-toxic} conversations reached a total approximately matching the number of \textit{toxic} conversations.
During this procedure, they identified 23 threads as toxic and 207 as non-toxic. The final set of 207 non-toxic threads constitutes our {\em \NTD}. The newly identified 23 toxic threads are combined with Ehsani et al's dataset of 175 toxic conversations. 

\subsection{\DTD} 
%\subsection{Derailment Points in Toxic Dataset}
% We focus on identifying the points at which conversations begin to derail in our toxic dataset. 
% Specifically, we observe that instances of incivility preceding toxic comments often indicate derailment, as they reflect a shift away from the original intent or tone of the discussion~\cite{zhang2018conversations}. 
Since we investigate on identifying derailed toxic threads, we further refined our dataset by identifying the points at which conversations begin to derail. 
Specifically, we observed that instances of incivility preceding the toxic comments often indicate derailment, as they signal a shift away from the original conversational intent~\cite{zhang2018conversations}. 
Given that Ehsani et al. previously annotated incivility in conversations containing TBDFs (see Table~\ref{tab:uncivil_features}), we hypothesize that any uncivil TBDF occurring before a toxic comment marks a potential derailment point. However, our investigation revealed that derailment can also arise outside of these TBDF categories, particularly in the presence of pronounced negative politeness strategies (e.g., \textit{`I’m terribly sorry, but'}, \textit{`Would you mind'})~\cite{danescu2013computational}. Therefore, we annotate the derailment points following the definitions provided in Section~\ref{sec:example} and use TBDFs as a guide whenever they were available.

% Some, albeit relatively few, toxic threads do not exhibit derailment; instead, in these conversations toxicity occurs suddenly and unexpectedly. Since our research aims to identify derailed toxic threads, we further refined our dataset by annotating whether derailment occurred before the appearance of toxic comments. While Ehsani et al.'s TBDF annotation indicated potential derailment points, we observed that their annotation didn't capture all possible instances, as their dataset is focused on specific categories of incivility, focusing on TBDFs only.  Derailment can occur beyond those TBDF categories, particularly when strong negative politeness~\cite{danescu2013computational} (e.g., \textit{`I’m terribly sorry, but'}, \textit{`Would you mind'}) are present~\cite{zhang2018conversations}. 

Two of the paper's authors reviewed the 198 toxic threads (175 from Ehsani et al.'s heated conversations dataset and 23 obtained while creating the {\NTD}) to identify possible derailment points preceding toxic comments. Their inter-annotator agreement reached 0.914 (Cohen's Kappa). 
The annotators determined that 34/198 threads exhibited "sudden toxicity", i.e., toxic threads do not exhibit derailment; instead, in these conversations toxicity occurs suddenly and unexpectedly.
% , i.e., toxic comments without prior derailment, which 
We excluded them from our analysis. 
The annotators discussed where they disagreed and resolved the disagreements. In 88 cases, they found multiple derailment points before the toxic comment. In 71 cases, annotators found exactly 1 derailment point. We also excluded 5 cases where the annotators could not agree whether there existed a derailment point or not. Finally, we retained 159 toxic threads (142 from Ehsani et al.'s and 17 newly identified threads from  Section \ref{nontoxic-dataset}) with 382 derailment points, which form our {\em \DTD}.

\change{
\subsection{Raman et al.'s Toxicity Dataset}\label{dataset_raman}
Since we propose a new approach in conversational derailment detection in SE domain, we experiment on evaluating the generalization of our proposed methodology. 
In order to evaluate that, we use the manually annotated toxicity dataset released by Raman et al.~\cite{raman2020stress}. It spans GitHub issue threads from 2012–2018, containing 167 toxic and 444 non-toxic threads, identified from locked or flagged discussions. However, there were only 314 threads available with comment-level annotations; since our evaluation setup (See more: Section~\ref{setup}) requires identifying the exact location of the first toxic comment, we exclude those without such detail. We also filtered out six threads that began with toxic comments, 308 threads remain (65 toxic, 243 non-toxic), making the dataset notably more imbalanced than our own. There is no overlap between this dataset and ours.
}

\begin{table*}[tb]
\caption{
\change{
Lexical cues in derailment point comments. 
Statistical significance (Chi-square test) between derailment and regular comments is indicated by {\em *} ($\alpha = 0.05$).
}}
    \centering
    \small
    \begin{tabular}{|l|c|c|c|c|c|}
        \hline
    \multirow{2}{*}{Linguistic Features} & \multicolumn{3}{c|}{Comment Type (\%)} & \multirow{2}{*}{$p$-value} & \multirow{2}{*}{Cramer’s $V$} \\ \cline{2-4}
    & Derail ($n=382$) & Toxic ($n=159$) & Regular ($n=1{,}371$) & & \\ \thickhline
        Second Person Pronouns & 60.7\% & 75.5\% & 43.9\% & $<0.0001^{*}$ & 0.127 \\ \hline
        WH Question Words & 57.1\% & 59.7\% & 43.9\% & $<0.0001^{*}$  & 0.104 \\ \hline
        Negation terms & 70.2\% & 71.1\% & 55.3\% & $<0.0001^{*}$  & 0.132 \\ \hline
        Reasoning terms & 70.4\% & 70.4\% & 61.4\% & $<0.0001^{*}$  & 0.055 \\ \hline
        Emphasis terms & 53.4\% & 59.7\% & 42.5\% & $<0.0001^{*}$  & 0.123 \\ \hline
        Communication Verbs & 33.5\% & 36.5\% & 24.9\% & $<0.0001^{*}$  & 0.133 \\ \hline
    \end{tabular}
    \noindent\parbox{\textwidth}
{\footnotesize\raggedright
   
    }
    \label{tab:discourses}
\end{table*}

\section{RQ1: What are the characteristics and patterns of conversational derailment in GitHub discussions? }
\label{properties_derailed_conversation}

Understanding how conversations derail on GitHub is fundamental to developing effective prediction systems.
Relative to conversations on other online platforms such as Reddit and Wikipedia, GitHub's technical conversations are distinct in the way that they exhibit toxicity~\cite{miller2022did}. 
Research also noted that diffient communities exhibit different characteristics in toxic communication patterns~\cite{song-etal-2025-echoes}.
Therefore, it stands to reason that conversational derailment on this platform may also be distinct. 
In this RQ, we empirically examine how conversational derailment manifests on GitHub. 
Specifically, we investigate: 1) the dynamics of conversational derailment by examining its timing and distance from the thread’s start, 2) the linguistic signals preceding derailment, 3) the presence of uncivil tones, and 4) common triggers that lead discussions off track.

We base our analysis on the \DTD. While limited in size and predominantly sourced from \textit{locked as heated} GitHub issues, this dataset is sufficient to observe derailment patterns. Locked issues are not a concern, as the locking occurs after the derailment and toxicity have occurred, and the locking mechanism does not affect the communication pattern. In addition, for the empirical analysis in this section, as our focus is on the interaction patterns leading up to toxicity, for any discussion thread, we exclude the comments  that occurred after the first toxic comment.

\subsection{Timing and Distance to Derailment Points} 
We calculate the median number of comments from the conversational first derailment point to the first toxic comment for each conversation thread in \DTD. 
\change{
In our toxic threads dataset, in terms of total comments, the median comment count 11, and mean comment count 17.6. However, the median first toxic comment occurrence position is 8 and mean is 12.03. And median first derailed comment occurrence position 4 and mean is 5.92. While, the median \textit{distance} is 3 comments and a mean \textit{distance} is 6.10.}
The close proximity between derailment and toxic comments suggests that once a thread derails, it is likely to directly devolve into toxicity. This aligns with Cheng et al.'s findings, which indicate that negative context and mood increase the likelihood of trolling behavior~\cite{cheng2017anyone}.

The timing of the first toxic comment relative to the derailment point provides additional insights. Considering a 8-hour workday, we observe about 46\% (73/159) of the of toxic comments occur within 8 hours of the first derailment comment~\cite{chang2019trouble} and about 64\% (102/159) occurs within 24 hours. This shows the importance of timely intervention. However, more than 25\% cases (40/159), the difference is more than 7 days, which indicates toxicity can also occur after a long period of derailment. This characteristic contrasts with platforms like Wikipedia, where a discussion is likely inactive if the last comment was added 2–3 days ago~\cite{schluger2022proactive}.

Note that in the few cases where there are multiple derailment point comments preceding the toxic comment, for this analysis, we considered the first derailment point in the conversation.

% \change{The timing of the first toxic comment relative to the derailment point provides additional insights. Considering, 8-hour workday, we observe more than half of the time (64\%, 102/159) of toxic comments occur within 8 hours of the derailment comment~\cite{chang2019trouble}. Of these 102, over 64.7\% (66/102) occurs within the first hour. This shows the importance of timely intervention. However, more than 14\% cases (23/159), the difference is more than 7 days, which indicates toxicity can also occur after a long period of derailment.}

\subsection{Linguistic Features}\label{sec:lang_features}
In Chang et al.'s study, participants noted that the easiest way to forecast conversational derailment is by analyzing user phrasing~\cite{chang2022thread}. Indicators include the use of direct address such as ‘you’ instead of generic terms like ‘all’ or ‘always’, as well as certain rhetorical postures or argumentative patterns can signal a conversation may be turning toxic~\cite{chang2022thread}.

We analyze potential language indicative of derailment in GitHub discussions. In the 382 derailment point comments in the \DTD, we sampled the 200 most frequent unigrams, excluding articles, particles, and common prepositions. 
\change{We intentionally retained negation and question terms because they are strong markers of argumentative or confrontational language in this context~\cite{strohm2018amplifiers, zhu2014negation, jiang2022nli, koshik2003wh}.}
Two annotators collaboratively categorized (see Table~\ref{tab:discourses}) the unigrams into linguistic groups using the card sorting method~\cite{schreier2012qualitative}. They met in person, discussed, and resolved differences, consulting a dictionary as needed. 
Based on these categories, we counted the frequency of each unigram in the derailment point comments after applying basic preprocessing steps (e.g., tokenization and lemmatization).

Table~\ref{tab:discourses} shows the percentages of occurrence in derailment points (382 count) along with the first toxic comments (159 count), and regular comments (non toxic and non derailment point comments); the total regular comments count were 1,371. 
We observed that in derailment points the elevated use of second person pronouns ('you', 'your', etc)~\cite{levy2022understanding}, negation terms (`not', `no', etc.), ``WH'' questions (`what', `why', `how', `where', etc.), reasoning terms (`because', `since', etc.), communication verbs (`say', `comment', `tell', etc.), and emphasis terms (`actually', `really', etc.) than general comments but lower than toxic comments. 

\change{As Table~\ref{tab:discourses} shows, all lexical differences between derailment and regular comments were statistically significant under a Chi-square test of independence~\cite{mchugh2013chi} ($\chi^2$, $p < 0.05$) after applying the Benjamini–Hochberg (BH) correction~\cite{benjamini1995controlling}. Effect sizes measured by Cramer’s $V$ (0.05–0.13) indicate small to moderate associations, confirming consistent but modest linguistic distinctions.}

Derailment point comments frequently combine structured reasoning with mild confrontation. Reasoning terms dominate (70.4\%), often paired with negation (70.2\%) and direct questioning (57.1\%), reflecting logical yet oppositional exchanges. Compared with regular comments, derailment points show +16.8\% more second-person targeting, +14.9\% more negation, and +13.2\% more questioning. Although they share similar reasoning intensity with toxic comments (70.4\% vs. 70.4\%), they contain less personal confrontation (60.7\% vs. 75.5\%) and reduced emphasis (53.4\% vs. 59.7\%). These findings suggest that derailment often emerges through argumentative yet non-abusive phrasing, where discussions shift from logical disagreement toward personal conflict.

% Our analysis reveals that derailment point comments often use structured arguments and moderately confrontational language. First, reasoning terms dominate derailment point comments, appearing in 70.4\% of cases (averaging 2.68 per comment), employing logical connectors to construct arguments within the discussion. Second, the argumentative language is supported by the occurrence of pervasive negation, with 70.2\% of derailment point comments containing contradiction terms (averaging 2.48 per comment) that systematically oppose and disagree with prior discourse. Finally, the confrontational nature emerges through direct targeting via second-person pronouns in 60.7\% of derailment point comments (1.98 average per comment) and challenging interrogation through WH-questions in 57.1\% of cases (1.53 average per comment). While not all WH-questions are inherently argumentative, their frequent use suggests a pattern of challenge-oriented engagement, particularly when combined with other confrontational markers. 

% When compared against regular non-derailed and non-toxic comments within the same toxic threads, derailment point comments demonstrate notable elevations: +16.8\% more second-person targeting, +14.9\% more negation, and +13.2\% more questioning than baseline discussion comments. While derailment point comments share comparable argumentative intensity with toxic comments (70.4\% vs. 70.4\% reasoning terms), they exhibit significantly less personal confrontation (60.7\% vs. 75.5\% second-person pronouns) and reduced emphatic language (53.4\% vs. 59.7\% emphasis terms). 

\subsection{Incivility TBDFs in Derailment Points}\label{sec:tbdf}
The tone of the comments is useful feature for proactive moderation. For example, moderators on Wikipedia assess tone to predict potential derailment~\cite{schluger2022proactive}.
Since the majority of the toxic conversations (142/159) in the {\DTD} came from Ehsani et al.'s dataset, the comments in these conversations already have incivility-related tones annotated. The Tone Bearing Discussion Features show the type of incivility at derailment point comments; we found 362 such annotated comments. For this analysis, we considered only those 362 derailed comments. Table~\ref{tab:tbdf} show the percentages of TBDFs in derailment comments and toxic comments.
% We measure the frequency of TBDFs in the derailment points and compare them to the TBDFs in the toxic comments in \DTD. 

The major uncivil TBDFs are: Bitter Frustration: 42.82\% (162/362), Impatience: 22.65\% (84/362), and Mocking: 9.94\% (39/362). 
For comparison, these same TBDFs occurred in toxic comments at the rates of 24.65\% (35/142), 9.15\% (13/142), and 11.97\% (17/142), respectively. Notably, while Insulting and Vulgarity are more prominent in toxic comments-- 25.35\% and 9.86\% respectively, they are less frequent in derailment points, occurring at 5.80\% and 2.49\%. This contrast indicates that certain forms of incivility such as \textit{Bitter Frustration} and \textit{Impatience} are more predictive of conversational derailment than direct toxicity. It also highlights a progression from subtle incivility to overt toxicity, and reinforces the importance of early signals such as frustration and impatience in anticipating derailment.

\begin{table}[tb]
\centering
%\footnotesize
\small
\caption{Top TBDF categories in derailment point comments (142 toxic threads from Ehsani et al.).}
\begin{tabular}{|l|c|c|}
\hline
{TBDF} & {Derail. Pt.} & {Toxic} \\
Category & Cmts. (362)  & Cmts. (142) \\ \thickhline
Bitter Frustration & 155 (42.82\%) & 35 (24.65\%) \\ \hline
Impatience         & 82 (22.65\%)  & 13 (9.15\%)  \\ \hline
Mocking            & 36 (9.94\%)   & 17 (11.97\%) \\ \hline
Insulting & 21 (5.80\%) & 36 (25.35\%) \\ 
% Irony & 16 (4.42\%) & 7 (4.93\%) \\
% Entitlement & 14 (3.87\%) & 6 (4.23\%) \\
% Vulgarity & 9 (2.49\%) & 14 (9.86\%) \\
% Threat & 3 (0.83\%) & 4 (2.82\%) \\
% Name-Calling & 2 (0.55\%) & 10 (7.04\%) \\
\hline
\end{tabular}
\label{tab:tbdf}
\end{table}

\subsection{Derailment Triggers}\label{sec:derailment_triggers}
Understanding what causes a conversation to derail can inform the development of effective early intervention strategies using automated, algorithmic approaches~\cite{schluger2022proactive, schaffner2024community}. 
Prior research in software engineering has identified potential triggers of toxicity in OSS discussions~\cite{ehsani2023exploring, ferreira2022how, miller2022did}. Ehsani et al.~\cite{ehsani2024incivility} proposed a guideline for annotating incivility triggers.

Building on their methodology, two authors independently annotated derailment triggers in our dataset, focusing on specific conversational or contextual elements that precipitated the initial shift. Although much of our toxic data overlaps with that of Ehsani et al., we chose to annotate derailment triggers with particular attention to the first derailment point in each conversation. This distinction was necessary because it was unclear from Ehsani et al.'s description whether their annotations considered the conversation holistically or targeted only the first uncivil comment when conversations started to go off-track.

The annotation achieved a Cohen's Kappa score of 0.84, indicating strong inter-annotator agreement. Disagreements were resolved through discussion to ensure full consensus. 
The most prevalent trigger was `Failed Use of Tool/Code or Error Messages' followed at 23.27\% (37/159), where tool difficulties or bug troubleshooting led to derailment. For example: \textit{``[CODE SNIPPET] ... What more proof do you need? That is everything."} The tension was caused here due to code error, which the user expressed in Frustrated tones. The conversation later evolved into toxicity. 
The second most prevalent derailment trigger was `Technical Disagreement' followed at 20.12\% (32/159), where tool difficulties or bug troubleshooting led to derailment. For instance: \textit{``[CODE SNIPPET] Ask yourself what **intention** it expresses. This is some kind of esoteric gibberish without reference to the subject area. [...]"}. In this case, the disagreements about method naming derails the conversation and later the conversation escalated to toxicity. Another major category was
`Communication Breakdown', which accounted for 16.98\% (27/159) of cases. This included misunderstandings, misinterpretations, typos, or language barriers causing perceived hostility. For example, \textit{``It is impolite to assume that each user opening an issue is stupid and lazy. Of course, I search the issue tracker. [...]"} Here, a misunderstanding between the commenters triggered conversation derailment.  

%\section{Conversation Derailment Prediction}
\section{RQ2: Can Large Language Models effectively predict conversational derailment on GitHub?}

% \textit{\textbf{RQ2:} Can Large Language Models effectively predict conversational derailment on GitHub?}

%This section describes our method of predicting conversational derailment on GitHub using LLMs. 
% In this section, we propose an LLM-based framework for predicting conversational derailment on GitHub. 
Building on RQ1's identification of derailment patterns, we investigate whether LLMs can effectively leverage conversation dynamics to predict derailment before toxicity occurs.
We draw inspiration from Hua et al., who demonstrated an automated derailment forecasting system by leveraging {\em Summaries of Conversation Dynamics} (SCD) for predictive modeling~\cite{hua2024did}.
SCDs offer a concise representation of a conversation's progression, characterizing the types of interactions that shaped its trajectory and forecasting future developments. Starting from Hua et al.’s original formulation, we incorporate insights from Section~\ref{properties_derailed_conversation} to reflect the conversational patterns observed to GitHub discussions.

\subsection{Baseline Models} 
We compare our SCD-based technique to two baselines:

\noindent
\textbf{CRAFT:} CRAFT is one of the earliest and best-known models for predicting conversational derailment~\cite{chang2019trouble}. The CRAFT tool is based on a neural network model. Since its inception in 2019, various other strategies for predicting conversational derailment have been explored, and CRAFT has been used as a baseline in these subsequent studies to evaluate the performance of newer approaches~\cite{altarawneh2023conversation, li2022multimodal, yuan2023conversation, kementchedjhieva2021dynamic}. 

\noindent
\textbf{Hua et al.'s approach}: Hua et al.~\cite{hua2024did} proposed that generating SCD can show the trajectory of the conversation and can be used to predict downstream conversational derailment task. They developed a few-shot procedural prompt where the LLM was provided with manually written SCD examples. Based on the prompt, they generated the SCD and predicted derailment as a downstream task.

% Since our approach adapts recent Hua et al.'s SCD-based technique for predicting conversation derailment ~\cite{hua2024did}, we also compare to their technique.

\subsection{LLM Prompt Design}\label{sec:prompt-design}
We design a two step LLM prompting procedure for GitHub derailment prediction:

\noindent
{\em Step 1 -- GitHub-Specific SCD Generation:} We convert the raw GitHub conversation into a high-level summary that captures interaction dynamics, emotional tone, and discourse strategies, excluding technical specifics.

\noindent
{\em Step 2 -- Derailment Prediction:} We estimate the probability of derailment based solely on the summary, using a simple scalar prediction prompt, 
\change{i.e., a prompt that instructs the LLM to produce a single numerical value, specifically, a probability between 0 and 1, representing the likelihood that a conversation will derail into toxicity.}
This prompt design separates the reasoning and classification stages, enabling us to build explainability into the pipeline and reducing the reasoning demands on the LLM.

\subsubsection{GitHub-Specific SCD Prompt}

We adapt Hua et al.~\cite{hua2024did}’s few-shot SCD prompt to GitHub’s conversational style and dynamics. Specifically, we replace general conversation examples with domain-specific interactions, such as discussions around pull request rejections or issue closures. Additionally, we instruct the model to ignore technical details like code snippets or file paths, and instead focus on conversational dynamics, i.e., how participants respond to one another, where misunderstandings arise, and how tone shifts over time. A typical summary generated using this GitHub-adapted SCD prompt is as follows:

% We explore several strategies to design the LLM prompt to predict conversational derailment on GitHub. 
% Firstly, we adapted Hua et al.~\cite{hua2024did}'s few-shot procedural SCD prompt for GitHub by specifically mentioning `GitHub conversation' and by providing examples based on GitHub conversations. An example of the SCD summary we provided in the prompt is as follows:

\smallskip
\begin{quote}
{\em
Multiple users debate reverting a recent PR. Speaker1 expresses strong opposition, referencing a previous incident involving similar code. Speaker2 challenges Speaker1’s framing and accuses them of misrepresenting past decisions. %[...]
Speaker3 supports Speaker2 and notes that Speaker1 had already raised this concern in another thread. Speaker1 becomes confrontational, citing a perceived pattern of dismissal. 
The conversation becomes increasingly heated as past interactions are used to question motives. The tone escalates, with limited signs of resolution.}
\end{quote}
\smallskip 

While this adaptation yielded clearer and more relevant summaries, we hypothesized that it may not yet be optimal. 
As Hua et al. developed SCD prompts targeting general-purpose conversations, it may not be most effective for the highly technical discussions found on GitHub. 
We explored whether decomposing the problem~\cite{khot2022decomposed} and
integrating the properties of GitHub derailed conversations, uncovered in Section~\ref{properties_derailed_conversation}, could yield better SCDs for predicting derailment on GitHub. Previous research shows that decomposing the prompts into incremental steps enhances the LLM's accuracy~\cite{khot2022decomposed, zhou2022least}.
Inspired by those studies, we adopted {\em Least-to-Most} (LtM) prompting strategy~\cite{zhou2022least}, and designed a step-wise summarization prompt that guides the model from high-level observations to detect smaller breakdowns in conversation patterns. This allows us to integrate our insights from Section~\ref{properties_derailed_conversation}. 
We incorporated the following key components into the LtM prompt (as Steps 3--6): 

\begin{enumerate}[itemsep=0pt,topsep=0pt,leftmargin=*]

\item \textbf{Individual Intentions (II)}, a feature we directly integrated from Hua et al.'s framework to analyze participants' motivations and goals. Morrill et al. similarly found that dialogue is shaped by intentions such as agreement, disagreement,  confrontation~\cite{morrill2024social}. Additionally, studies have shown that communication styles play a role in shaping interactions within GitHub discussions~\cite{batoun2023empirical, wang2023conversations}.

\item \textbf{Conversational Features (CF)} (e.g., questioning, rhetoric), where we adopted the categories established by Hua et al.~\cite{hua2024did}, which includes rhetorical questions, hedging, questioning logic, and other linguistic patterns. This approach aligns with our findings described in Section~\ref{sec:lang_features}; 

\item \textbf{Sentiment and Tonal Features (STF)}, which capture emotional dynamics and shifts throughout the discussion, enabling us to track how sentiment evolves before derailment occurs, as discussed in Section~\ref{sec:tbdf}; and

\item \textbf{Tension Triggers (TT)}, which identify potential catalysts for escalating conflict that serve as early indicators of possible derailment, based on our findings described in Section~\ref{sec:derailment_triggers};  

\end{enumerate}

We explicitly instruct the model to exclude technical content and focus on interactional dynamics. The summary is synthesized at the final step.

In developing the LtM prompt, we initially experimented with prompts that included explicit definitions for each reasoning step, for instance, describing categories of tension triggers, conversation strategies or specifying tonal cues like sarcasm or frustration. However, we observed that such definitions often introduced unnecessary verbosity and led to inconsistent outputs in SCD generation. 
In contrast, we found that a more concise, open-ended prompt, omitting explicit definitions, led to more consistent and focused outputs. Rather than prescribing rigid categories, this design allowed the model to apply its learned understanding of conversational structure. The final LtM prompt is as follows:

\begin{promptbox}{Least-to-Most (LtM) SCD Generator Prompt}
% \sf
{
\small
% \raggedright
% \palatino
\noindent
You are a skilled Conversation Analyst specializing in GitHub discussions. Your objective is to capture the conversation dynamics without getting caught in the technical details. \newline

\textbf{Your Analysis Method:} Follow these steps in order, building your understanding from basic patterns to complex dynamics: \newline

\textbf{Step 1:} Identify Main Elements: Identify the main elements by quickly scanning the conversation and pinpointing the key components or topics being discussed. \newline

\textbf{Step 2:} Enforce Exclusion Criteria\newline
Do NOT include:\newline
- Any technical claims, arguments, or explanations \newline
- Any code names, module names, or PR details \newline
- Any direct or indirect quotations \newline
- Any mention of what was being implemented or reviewed \newline
Example 1 (Excluded):
"The discussion addressed discrepancies in $evaluate\_admissions()$ outputs [...]."
\newline

\textbf{Step 3:} Note Individual Intentions\newline
Infer what each participant is aiming to achieve. Example: [...]
 \newline

\textbf{Step 4:} Identify Conversation Strategies\newline
Identify rhetorical or structural tactics used by each speaker. Example: [...] \newline

\textbf{Step 5:} Track Sentiment Evolution\newline
Describe how the tone shifts across the conversation. Use descriptors such as neutral, guarded, dismissive, confrontational, collaborative, defensive, escalating. Avoid inference.  [...] \newline

\textbf{Step 6:} Find Tension Triggers\newline
Detect explicit shifts in tone, rhythm, or interaction style that signal rising tension—only when clearly supported by observable evidence. Avoid inference. [...] \newline

\textbf{Step 7:} Synthesize Trajectory Summary \newline
Combine insights from steps 1-6 to create a concise trajectory summary. \newline
Remember: Your value lies in revealing the human dynamics beneath technical discussions. Focus on HOW people communicate, not WHAT technical points they make. \newline
Example: "Multiple users debate reverting a recent PR.[...]" \newline

\textbf{Your Task:} Using the above guideline, write the final trajectory summary for the following GitHub discussion transcript:
\newline

$<insert\ conversation\ transcript>$ \newline

Write only the final summary within double quotation.
}

\end{promptbox}

Here is an example SCD summary generated using this prompt:

\smallskip
\begin{quote}

{\em 
USER\_01 initiates a discussion about addressing deprecation warnings in a project, suggesting bumping the version number as a solution. 
USER\_02 and USER\_03 express reluctance to create a new major version, leading USER\_01 to interpret this as a lack of future planning for the project. %[...]
The conversation shifts to a more confrontational tone when USER\_02 suggests an alternative fix and sarcastically requests an apology from USER\_01, which escalates USER\_01's response into a threat of ceasing contributions and blocking participants. 
USER\_02 attempts to de-escalate by reiterating the availability of an alternative solution but maintains a guarded stance.
}
\end{quote}

\smallskip

% We observe that this SCD summary was more effective at capturing the emotional aspects and tone of the GitHub conversations, compared to our initial approaches. 

\subsubsection{Derailment Prediction Prompt}
We employ a follow-up prompt that asks the model to predict derailment probability (0-1 scale) based on the generated SCD summary:

\begin{promptbox}{Derailment Predictor Prompt}
{
\small
% \raggedright
% \palatino
You will be provided with a summary of a GitHub conversation trajectory. A Conversation Trajectory Summary captures the evolution of a discussion by identifying key elements, triggers of tension, conversational strategies, interaction patterns, sentiments, and tonal shifts among participants. It provides a concise overview of significant dynamics and tone changes without detailing the technical aspects or specific arguments.
\newline

Now based on the given GitHub conversation trajectory summary, what is the probability that the conversation may derail and become toxic on a scale of 0 to 1 (Toxicity is defined by “any rude, disrespectful, or unreasonable comment that is likely to make one leave a discussion.”)?
\newline

Conversation trajectory summary: $<insert\ summary>$
\newline

Write exactly one word: the probability rounded to two decimal places.
%\newline
%
% Do not write reasoning.
}

\end{promptbox}

This two-step prompting process allows us to first generate a comprehensive summary of the conversation and then use that summary to make a more informed prediction about the potential for toxicity.

\subsection{Experiment Setup}\label{setup}

\subsubsection{Large Language Models}
We conduct all of our experiments using two publicly available LLMs: Llama (\textit{llama-3.3-70B} version) and Qwen (\textit{qwen2.5:32b-instruct} version) model, as they are among the top-performing open-weight state-of-the-art LLMs at the time of conducting this study. We specifically chose freely available LLMs and avoided proprietary models (e.g, GPT-4 or Claude) to improve reproducibility and because the cost of paid models may make them impractical for real-world deployment.

We set the model temperature to 0 to minimize output variance and set a uniform a context window size of 32k (maximum length supported by Qwen 2.5 version). For each toxic conversation in our dataset, we provide all the comments up to, but excluding, the first toxic comment. 
\change{
Formally, let $C = \{c_{1}, c_{2}, \ldots, c_{n}\}$ be the ordered set of comments in a GitHub conversation, and let $c_{t}$ be the first toxic comment such that:  
$t = \min \{\, i \mid c_{i} \text{ is labeled toxic} \,\}$,
then, the model input is:  
$$C' = \{c_{1}, c_{2}, \ldots, c_{t-1}\}.$$
Thus, the model only sees conversation context before toxicity emerges. It does not have access to the toxic comments themselves (and any comments afterwards). This formulation follows the forecasting prediction setup used by Chang et al.~\cite{chang2019trouble}.
}

Whenever a conversation exceeded the context window limit, we truncated it by removing utterances from the beginning until the total length fell within the allowable range.

\subsubsection{Metrics}
We compare three popular metrics: %Precision, Recall and F1-score.

\begin{itemize} %[itemsep=0pt,left=0pt]
    \item  \textbf{Precision} refers to the proportion of true positive observations among all the predicted positive observations:\\ $\text{Precision} = \frac{\text{True Positive}}{\text{True Positive + False Positive}}.$
    \item \textbf{Recall} represents the proportion of true positive observations out of all actual positive observations:\\ $\text{Recall} = \frac{\text{True Positive}}{\text{True Positive + False Negative}}.$
    \item The \textbf{F1-score} is the harmonic mean of Precision and Recall: $\text{F1-score} = 2 * \frac{\text{Precision * Recall}}{\text{Precision + Recall}}.$
\end{itemize}

\begin{table}[tb]
    \caption{Derailment prediction results for different models on Derailed Dataset and Non-Toxic Dataset. (SCD = Summaries of Conversation Dynamics, $\theta$ = $Threshold$).}
    \centering
    %\scriptsize
    \small
    \begin{tabular}{|l|c|c|c|c|c|}
        \hline
        \multirow{1}{*}{Model} 
        & \multirow{1}{3em}{\centering{Strategy}} 
        & \multirow{1}{3em}{\centering $\theta$ ($\geq$)}
        & \multirow{1}{3.5em}{\centering{Precision}} 
        & \multirow{1}{2.5em}{\centering{Recall}}
        & F1\\
        \thickhline
        \multirow{3}{4em}{{CRAFT \cite{chang2019trouble}}}
        & - & 0.1 & 0.419 & 0.937 & 0.579 \\ 
        & - & 0.3 & 0.425 &	0.912 &	0.580 \\ 
        & - & 0.5 & 0.585 & 0.522 & 0.551 \\
        & - & 0.7 & 0.764 & 0.346 & 0.476 \\
\thickhline

\multirow{8}{4em}{{Qwen 2.5:32B Instruct}}
    & \multirow{4}{4em}{{Hua et al. FewShot SCD~\cite{hua2024did}}}
       & 0.1 & 0.440 & 1.000 & 0.612 \\ 
        & & 0.3 & 0.980 & 0.604 & 0.747 \\ 
        & & 0.5 & 1.000 & 0.258 & 0.410 \\ 
        & & 0.7 & 1.000 & 0.164 & 0.281 \\ 
    \cline{2-6}
    & \multirow{4}{4em}{{Least-to-Most SCD}}
        & 0.1 & 0.443 & 1.000 & 0.614 \\ 
        & & 0.3 & 0.945 & 0.862 & \textbf{0.901} \\ 
        & & 0.5 & 0.987 & 0.478 & 0.644 \\ 
        & & 0.7 & 0.981 & 0.327 & 0.491 \\ 
    \thickhline
        
\multirow{8}{4em}{{Llama 3.3:70B}}
    & \multirow{4}{4em}{{Hua et al. FewShot SCD~\cite{hua2024did}}}
          & 0.1 & 0.750 & 0.981 & 0.850 \\ 
        & & 0.3 & 0.913 & 0.723 & 0.807 \\ 
        & & 0.5 & 0.929 & 0.572 & 0.708 \\ 
        & & 0.7 & 0.983 & 0.358 & 0.525 \\ 
    \cline{2-6}
    & \multirow{4}{4em}{{Least-to-Most SCD}}
        & 0.1 & 0.702 & 0.962 & 0.812 \\ 
        & & 0.3 & 0.890 & 0.818 & \textbf{0.852}\\ 
        & & 0.5 & 0.932 & 0.686 & 0.790 \\ 
        & & 0.7 & 0.975 & 0.491 & 0.653 \\ 
    \hline
    \end{tabular}
    \label{tab:models_accuracy}
\end{table}
\renewcommand{\arraystretch}{1.0} % Default value: 1

\subsection{Results and Discussion}
\label{results}
We conduct experiments using our dataset of 159 derailed toxic threads and 207 non-toxic threads, a total of 367 data points. Of these, none of the threads exceeded the context window limit. Chang et al. showed that the prediction decision threshold (i.e., probability cutoffs) can vary widely for different datasets in the conversational derailment task~\cite{chang2019trouble}. Since we asked the LLMs to provide a prediction score of derailment between 0 to 1, we include results from different thresholds, $\theta \in \{0.1, \ 0.3, \ 0.5, \ 0.7\}$. The results of this experiment are shown in Table~\ref{tab:models_accuracy}.  

Our Least-to-Most SCD prompting strategy demonstrates strong performance across both models, achieving F1-scores of 0.901 (Qwen) and 0.852 (Llama) at the 0.3 threshold. The approach maintains balanced performance with high precision (0.945 for Qwen, 0.890 for Llama) while preserving good recall (0.862 for Qwen, 0.818 for Llama). Compared to the two baselines, our method significantly outperforms CRAFT, which consistently underperformed across all thresholds with best F1-score at 0.580. Against Hua et al.'s SCD prompting strategy, our approach shows superior results at most thresholds, particularly excelling at the practical 0.3-0.7 range where balanced precision-recall trade-offs are crucial~\cite{chang2022thread}.

For the Qwen model, the Least-to-Most strategy achieved the highest F1-scores across all threshold values. A similar trend was observed for the Llama model, with the exception of the 0.1 threshold where Hua et al.'s strategy performed better due to very high recall (156/159 correct predictions) but at the cost of much lower precision. This trade-off highlights an important consideration: while high recall captures most toxic instances, low precision leads to excessive false positives. In GitHub repositories where non-toxic conversations vastly outnumber toxic ones, maintaining high precision is critical to avoid unnecessary false alarms~\cite{raman2020stress}. However, the high precision should not come at the cost of low precision as Chang et al. noted that balancing precision and recall is crucial, as too many false positives reduce a tool's helpfulness while too many false negatives reduce effectiveness~\cite{chang2022thread}. At higher thresholds ($\geq$ 0.3), our Least-to-Most strategy ensures good F1-scores with high precision, making it practically viable for deployment.

Based on these results, we envision a threshold-based intervention strategy to mitigate toxicity: higher thresholds (e.g., $\theta > 0.7)$ could alert moderators to review flagged content, while lower thresholds (e.g., $\theta = 0.3 \ to \ 0.7)$ could trigger automated bots to issue reminders promoting civil discourse.

\subsection{Error Analysis}~\label{sec:error} 

To better understand the performance and limitations of the model and datasets,  we conduct an error analysis. We limited the error analysis to the best-performing configuration, i.e., the LtM SCD prompting strategy on the Qwen model at 0.3 threshold. 
\change{
We conducted an open coding process to analyze the errors~\cite{williams2019art}. Two authors independently reviewed all misclassified cases and assigned preliminary labels to them. They then met to compare their labels, resolve disagreements, and refine the categories through iterative discussion. This process was repeated until complete consensus was achieved.
}

There were two types of error categories:
1) 8 cases where the model predicted non-toxic conversations as derailing; and
2) 22 cases where the model predicted derailed toxic conversations as non-derailing. Two authors reviewed the conversations, examined the generated SCD, and determined the most likely reason for the error. They finalized the error categories using open coding, and further improved them thorough discussion and applying axial coding~\cite{charmaz2006constructing}, with some cases belonging to more than one category. 

 In the 8 false positives, the main issues were the \textit{model overestimating tension} in otherwise civil exchanges (3 cases), and situations where the \textit{SCD was accurate but the predictor misjudged} the tone’s seriousness with respect to toxicity (3 cases). Among the 22 false negatives, most errors were due to \textit{missing or underestimating subtle toxic signals} like frustration (10 cases), \textit{failure to detect sarcasm or nuanced tones} (3 cases), \textit{accurate SCDs with flawed predictor judgment} (3 cases), and cases where \textit{toxicity occurred long after derailment} (3 cases), reducing the perceived severity.
\change{
To illustrate, consider the following SCD where the model overestimated the seriousness of the tone as confrontational when the commentators were sharing their perspective, ``\textit{Participants discuss changes to game mechanics involving Repair Facility and Heavy Repair Turrets (HRT). [... ] The conversation shifts from neutral to confrontational as participants assert their viewpoints and express dissatisfaction with proposed changes.}''. In reality, participants were simply exchanging perspectives without hostility. In another case, the model missed sarcasm about image quality: ``\textit{Two users discuss an image and a JSON file issue. [...] reiterating concerns about image quality, the tone remains critical but not confrontational, focusing on clarity and quality standards.}'' Here, subtle humor and sarcasm were misinterpreted as neutral critique, highlighting the model’s difficulty with nuanced tones.
}
These findings highlight specific weaknesses in both tone interpretation and the alignment between SCD generation and downstream prediction.

\begin{table}[tb]
\centering
%\scriptsize
\small
\caption{Ablation Study on our Least-to-Most (LtM) Derailment Prediction Strategy 
using Qwen and $\theta$ = $0.3$.}
\begin{tabular}{|l|c|c|c|c|}
\hline
\textbf{Ablation} & \textbf{Precision} & \textbf{Recall}& \textbf{F1} & \textbf{$p-value$}  \\
&   ($\Delta$) & ($\Delta$) &  ($\Delta$) & (Significant) \\
\hline

No II &   0.962 (+1.8\%) &   0.792 (-8.1\%) &  0.869 (-3.6\%) &      0.2153 ($\times$) \\
No CF & 0.941 (-0.4\%) &   0.805 (-6.6\%) &  0.868 (-3.7\%) &      0.1628  ($\times$) \\
No STF & 0.898 (-5.0\%) &  0.774 (-10.2\%) &  0.831 (-7.8\%) &      0.0032 ($\checkmark$) \\
No TT & 0.906 (-4.1\%) &   0.786 (-8.8\%) &  0.842 (-6.5\%) &      0.0190 ($\checkmark$) \\
\hline
Full LtM & 0.945 & 0.861 & 0.901 & -- \\
\hline
\end{tabular}
\label{tab:ablation}
\end{table}

\subsection{Ablation Study}
To better understand the contribution of individual semantic components in our summarization prompt, we conducted an ablation study focused on our Least-to-Most (LtM) derailment prediction. 

As introduced in Section~\ref{sec:prompt-design}, the LtM prompt integrates multiple high-level conversational features: Sentiment and Tonal Features (STF), Individual Intentions (II), Conversation Features (CF), and Tention Triggers (TT). Through the ablation study, we aim to quantify the effect of removing specific components in shaping the predictive utility of the summaries. We evaluate ablations exclusively on the Qwen model at $\theta = 0.3$ threshold because it achieved the best performance in our experiments.

\subsubsection{Prompt Modifications Per Component}

Each ablation removes one semantic component from the original LtM prompt:

\begin{itemize} %[itemsep=0pt,topsep=0pt,leftmargin=*]
\item Removed \textbf{II} ({LtM Prompt Step 3:} \emph{Note Individual Intentions}); renumbered subsequent steps.
\item Removed \textbf{CF} ({LtM Prompt Step 4:} \emph{Identify Conversation Strategies}); renumbered subsequent steps.
\item Removed \textbf{STF} ({LtM Prompt Step 5:} \emph{Track Sentiment Evolution}); renumbered subsequent steps.
\item Removed \textbf{TT} ({LtM Prompt Step 6:} \emph{Find Tension Triggers}); renumbered subsequent steps.
\end{itemize}

All other steps and examples were preserved to maintain structural consistency. The goal was to generate SCDs comparable to those from the baseline strategy. The full ablation prompts are included in the replication package. In addition to precision, recall and F1-score, we also perform statistical significance test by employing McNemar’s test~\cite{mcnemar1947note}. To control for multiple comparisons, we applied the Benjamini-Hochberg (BH) correction~\cite{benjamini1995controlling}.

\subsubsection{Results}
Table~\ref{tab:ablation} presents the precision, recall, F1-score, and BH corrected $p$-values from McNemar’s test for each feature ablation. McNemar’s test evaluates whether differences in model predictions are statistically significant under paired comparisons. A $p$-value below 0.05 denotes a statistically significant change in predictions.

Ablating the \textbf{STF} (sentiment and tone features) component results in the largest F1-score reduction (-7.8\%), primarily due to a -10.2\% drop in recall and a -5.0\% decrease in precision. This change is statistically significant ($p = 0.0032$), indicating that STF features play a critical role in recall-oriented classification performance.

Removing \textbf{TT} (tension triggers) leads to a -6.5\% F1 decline, driven by an -8.8\% reduction in recall and a -4.1\% drop in precision. The prediction shift is statistically significant ($p = 0.0190$), confirming the contribution of trigger features to predictive performance.

Ablating \textbf{II} results in a smaller F1 decrease (-3.6\%), with a minor gain in precision (+1.8\%) and a larger drop in recall (-8.1\%). The BH-corrected $p$-value ($p = 0.2153$) does not indicate statistical significance. Likewise, removal of the \textbf{CF} feature yields a $-3.7\%$ F1 decrease, with minimal change in precision (-0.4\%) and a -6.6\% drop in recall; this change is also not statistically significant.

Overall, the results show that STF and TT features have the strongest and statistically significant impacts on out LtM prompted model performance, particularly in preserving coverage (recall).

\section{RQ3: To what extent does the proposed LLM-based derailment prediction approach generalize to independent GitHub datasets?}\label{different_dataset}

\change{
In order to validate our approach's generalizability, we evaluate it on Raman et al.'s~\cite{raman2020stress} publicly available dataset~\cite{raman-toxicity-detector}, which is independent of our curated data and follows a different annotation procedure.
}

\subsection{Dataset and Experimental Setup}
% 
% The dataset spans GitHub issue threads from 2012 to 2018, with 167 labeled as toxic and the remaining as non-toxic. The toxic threads were sourced from (i) heated locked issues or (ii) comments containing the word \textit{`attitude.'} The dataset further includes 444 non-toxic threads.  
% However, we found that only 314 threads have comment-level annotations available in their replication package, while the rest data are provided at the conversation level. 
% Since \change{as formulated in Section~\ref{setup}, evaluating the conversational derailment prediction requires comment-level annotation to find the exact toxic comment location}, we discarded the threads where comment level data were not available. 
% Of the remaining 314 conversations, six threads began with a toxic comment, which we excluded in accordance with our experimental criteria. This filtering resulted in a final set of 308 GitHub threads. Of these, 65 were labeled as toxic and 243 as non-toxic, making \textit{the dataset significantly more imbalanced than our own}. There is no overlap between this dataset and our own curated dataset. 

\change{
As introduced in Section \ref{dataset_raman}, the dataset comprises 168 toxic and 444 non-toxic threads. However, we found 314 threads have comment-level annotations available in their replication package.
Since, evaluating the conversational derailment prediction requires comment-level annotation to find the exact toxic comment location, we filtered those 314 conversations.
We further filter 6 conversations where toxicity observed at first comment. Therefore, we end up with 308 GitHub issue threads (65 toxic, 243 non-toxic).
We apply the same preprocessing and modeling pipeline described in Section \ref{setup} to ensure comparability across datasets.
}
Note that the dataset may possibly include sudden toxic conversations. We have not verified them manually.

As before, we evaluated the two prompt based techniques on this dataset using the Llama (\textit{Llama-3.3:70B-3.3:70B} version) and Qwen (\textit{Qwen-2.5:32B-Instruct} version) model, setting the thresholds: $\theta$ $\in$ $\{0.1, 0.3, 0.5, 0.7\}$. We report Precision, Recall, and F1-score as before for each setting.

\subsection{Results and Discussion}
The results are presented in Table~\ref{tab:experiment_raman}.  The Qwen model outperformed Llama on this benchmark, consistent with trends observed in our curated dataset. The LtM SCD strategy achieved the highest F1-score of 0.797 at a threshold of 0.3. For the Llama model, the same strategy yielded the best F1-score of 0.776, at a threshold of 0.5.

These findings reaffirm the effectiveness of the LtM prompting strategy over the baseline Hua et al. SCD few-shot prompt in detecting early conversational derailment on GitHub. While Qwen produced stronger results overall, Llama exhibited more stable performance across thresholds. Additionally, lower thresholds increased recall at the expense of precision, highlighting the importance of threshold calibration in real-world moderation settings.

\begin{table}[tb]
\caption{Derailment prediction results for different models on Raman et al.'s dataset.}
%\footnotesize
\small
\centering
\begin{tabular}{|l|c|c|c|c|c|}  
    \hline  
    Model
    & Strategy
    & $\theta$
    & Precision
    & Recall  
    & F1 \\  
    \thickhline  
    
    \multirow{8}{4em}{Qwen} & \multirow{4}{4em}{Hua et al. FewShot SCD}
          & 0.1 & 0.280 & 1.000 & 0.438 \\ 
        & & 0.3 & 0.807 & 0.708 & 0.754 \\ 
        & & 0.5 & 0.875 & 0.431 & 0.577 \\ 
        & & 0.7 & 0.955 & 0.323 & 0.483 \\ 
    \cline{2-6}
    
    & \multirow{4}{4em}{Least-to-Most Strategy}
    & 0.1 & 0.236 & 1.000 & 0.382 \\ 
    & & 0.3 & 0.753 & 0.846 & \textbf{0.797} \\ 
    & & 0.5 & 0.816 & 0.615 & 0.702 \\ 
    & & 0.7 & 0.857 & 0.369 & 0.516 \\ 
    \thickhline

    \multirow{8}{4em}{Llama} & \multirow{4}{4em}{Hua et al. FewShot SCD}
      & 0.1 & 0.594 & 0.877 & 0.708 \\ 
    & & 0.3 & 0.746 & 0.723 & 0.734 \\ 
    & & 0.5 & 0.804 & 0.631 & 0.707 \\ 
    & & 0.7 & 0.853 & 0.446 & 0.586 \\ 
    \cline{2-6}
    
    & \multirow{4}{4em}{Least-to-Most Strategy}
      & 0.1 & 0.513 & 0.908 & 0.656 \\ 
    & & 0.3 & 0.659 & 0.831 & 0.735 \\ 
    & & 0.5 & 0.754 & 0.800 & \textbf{0.776} \\ 
    & & 0.7 & 0.804 & 0.692 & 0.744 \\ 
    \hline

\end{tabular}  
\label{tab:experiment_raman}  
\end{table}
\renewcommand{\arraystretch}{1.0}

% \change{MAYBE ANOTHER EXPERIMENT WITH A COMPLETELY NEW UNLABELED DATASET?}

\section{Recommendations}

The results indicate several practical directions for improving proactive moderation in open-source communities, particularly on GitHub. Based on empirical observations and model behavior across datasets, \change{we outline recommendations for two groups: (1) researchers studying conversational derailment, and (2) GitHub maintainers involved in community moderation.}

\noindent
\textbf{Recommendations for GitHub Maintainers.}
GitHub repository maintainers can integrate LLM-based early warning systems into existing moderation pipelines. Summarization-driven derailment detection provides a lightweight mechanism to flag discussions that may require attention, enabling timely intervention without exhaustive manual review.

Intervention thresholds should be calibrated according to moderation goals. Higher thresholds (e.g., $>$ 0.7) are suited for passive alerts aimed at de-escalation~\cite{qiu2023climate}, while intermediate thresholds (e.g., 0.3–0.7) can trigger automated reminders that act as conversational mediators~\cite{schluger2022proactive}. 
This can be integrating within existing GitHub's infrastructure. Building on GitHub’s current tagging options (e.g., `abuse', `off-topic', `resolved') for ~\cite{githubGraphQLMinimizable}, a dedicated derailment' tag could streamline moderator review and response.
% \change{This can be integrating within existing GitHub's infrastructure. Building on GitHub’s \texttt{Minimizable} functionality~\cite{githubGraphQLMinimizable}, implemented across entities such as \texttt{Issue Comment}, \texttt{Pull Request Review}, and \texttt{Discussion Comment}, which supports reasons like \textit{abuse}, \textit{off-topic}, \textit{outdated}, \textit{resolved}, \textit{duplicate}, and \textit{spam}, we suggest introducing an additional reason type for \textit{derailment} to streamline moderator review and response. 
% }
% Figure~\ref{fig:github_adoption} illustrates a possible extension of GitHub’s minimization interface.

\change{
Maintainers could incorporate the model’s SCDs into issue and pull request dashboards to identify threads showing early signs of tension. 
Integration frequency should align with repository activity and cost constraints. Since 64\% of toxic exhanges occur within 24 hours of derailment, higher-frequency runs (e.g., hourly) are most useful for new or rapidly evolving threads, whereas running the model after each new comment suffices for slower discussions. The SCDs provide interpretable summaries that can help maintainers make informed moderation decisions.
}

% \begin{figure}[t]
% \centering
% \includegraphics[width=\linewidth]{images/github_adoption.pdf}
% \caption{Proposed addition of a ‘Derailment’ reason to the GitHub Minimizable interface for structured moderator feedback.}
% \label{fig:github_adoption}
% \end{figure}

\noindent
\textbf{Recommendations for Researchers.}
Future research should build upon the current study by refining prompt designs to capture conversational signals of derailment more consistently across contexts. While this work has demonstrated the value of modeling tonal shifts, tension triggers, and sentiment trajectories, further efforts are needed to generalize these features across diverse OSS communities, languages, and moderation norms. 
Refinement should also aim to reduce prompt sensitivity and improve reproducibility of LLM-based summarization across datasets.

Developing standardized benchmarks across multiple platforms, such as GitHub and BugZilla, would enable consistent evaluation and cross-comparison of models. These benchmarks should include annotated derailment points, incivility categories, and moderation outcomes.

\change{
Efficiency and scalability remain key challenges. Incrementally updating summaries with new comments, rather than reprocessing entire threads, could allow near–real-time assessment with reduced computational cost. Beyond accuracy, researchers should evaluate latency, resource trade-offs, and feedback dynamics between predictive models and human moderators.
}

\change{
Transparency and explainability should remain research priorities. Future work should further refine how SCD convey rationales for tension and escalation, improving trust and accountability~\cite{schluger2022proactive}.
}

% Effective prompt design should prioritize conversational features such as tonal shifts and tension triggers over technical content, as these elements more reliably signal potential derailment. 

% The field would likely benefit from the development of standardized benchmarks across multiple OSS platforms, encompassing annotated derailment points, incivility/toxicity types, and moderation outcomes. 

% Finally, to foster trust and accountability in automated systems, transparency must be emphasized. Schluger et al. noted that several moderators in Wikipedia remarked that they wanted to know \textit{`why those conversations might derail'}~\cite{schluger2022proactive}. The generated SCDs offer an opportunity for more transparent and explainable moderation practices, providing clear justifications for interventions and allowing both users and moderators to understand the basis of predictive outcomes.

\section{Related Work}

Our work builds upon and extends previous research in two main areas: toxicity analysis in software engineering and conversational derailment prediction.

\noindent \textbf{Toxicity analysis in SE artifacts.}
A large body of work has explored negative communication in developer-facing platforms such as GitHub, Stack Overflow, and code review tools. Studies have examined incivility~\cite{ferreira2024incivility, ferreira2021shut, ehsani2024incivility}, emotional tone~\cite{imran2022data, imran2024uncovering, novielli2020can, qiu2022detecting}, toxicity~\cite{raman2020stress, sarker2020benchmark, sarker2025landscape, cohen2021contextualizing} and their effects on contributor engagement~\cite{miller2022did, imran2025silent, tian2024analyzing, rahman2024words}. 

Several investigations have examined the consequences of toxicity. For example, toxic interactions have been linked to developer stress and increased dropout rates~\cite{raman2020stress}. Other work has explored moderation strategies, the role of bots, and the benefits of early detection tools in mitigating toxic dynamics~\cite{miller2022did, hsieh2023nip}. Additionally, analysis of locked discussions has provided insights into recurring patterns of incivility and contributed key datasets~\cite{ferreira2022how, ehsani2024incivility}. 

Researchers have developed tools to automate toxicity detection in software engineering~\cite{raman2020stress, sarker2023automated, mishra2024exploring}. However, all those tools are \textbf{post-hoc}, addressing toxicity only after it has occurred. Our work builds on this line by proposing a \textit{proactive detection strategy}, shifting from retrospective classification to early forecasting of derailment events that precede toxicity.

\noindent \textbf{Conversation derailment.}
Predictive studies of derailment have largely centered around Wikipedia and Reddit~\cite{zhang2018conversations, levy2022understanding, bao2021conversations, chang2019trouble, li2022multimodal, hua2024did}. The CRAFT model~\cite{chang2019trouble} pioneered early detection of online toxicity via linguistic and structural features. Hua et al.~\cite{hua2024did} introduced the concept of \textit{Summaries of Conversation Dynamics (SCD) }for forecasting conversational trajectory, a foundation we extend using GitHub-specific dynamics. 

Recent work also investigates hierarchical transformers, neural network, user behavior modeling, and contextual features such as reply structure and edit markers~\cite{kementchedjhieva2021dynamic, li2022multimodal, de2021beg, hessel2019something, anuchitanukul2022revisiting, yuan2023conversation, altarawneh2023conversation}. However, these approaches often lack domain adaptation for technical forums like GitHub, where toxicity emerges through more subtle signals like entitlement or miscommunication~\cite{miller2022did, hsieh2023nip}.

Our contribution lies in integrating LLM-driven strategy with structured prompting tailored to a technical platform like GitHub’s, providing both performance and explainability in derailment forecasting.

\section{Threats to Validity}

We note potential threats to the validity of our study in the following categories: construct validity, internal validity, and external validity.

\noindent \textbf{Construct validity.} 
Construct validity concerns whether our methodology accurately captures the concept of derailment as a precursor to toxicity. A key limitation is that not all toxic behavior emerges gradually; some instances arise abruptly without prior conversational signals, making them undetectable by our approach. Additionally, derailment is annotated based on human judgment, which may vary across annotators. LLM-generated SCDs can also introduce hallucinations or overlook subtle shifts in tone. To mitigate these risks, we used detailed annotation guidelines, ensured high inter-annotator agreement, and conducted error analyses to identify common misclassifications.

\smallskip
\noindent \textbf{Internal validity.} 
Internal validity relates to whether the results are attributable to our method rather than uncontrolled factors. Our use of LLMs introduces variability due to stochastic generation and prompt sensitivity; small changes in input can affect model outputs. Moreover, although we followed a structured annotation process, human error or bias may still influence labeling consistency. We addressed these issues by employing a model-in-the-loop framework, cross-checking annotations, and designing prompts systematically to minimize ambiguity.

\smallskip
\noindent \textbf{External validity.} 
External validity reflects the generalization of our findings beyond the specific dataset and setting. Our dataset is limited to GitHub issue threads and may not generalize to other platforms such as JIRA, GitLab, or non-OSS communities, which have different conversational norms. Additionally, the curated dataset is relatively small, potentially limiting applicability across all GitHub communities. 
\change{
Furthermore, because many public GitHub discussions are part of the web-scale data used in LLM pretraining, it is possible that some threads in our dataset overlap with or resemble the model’s pretraining data. This potential exposure may slightly inflate performance estimates, underscoring the need to validate the framework on data from unseen platforms or domains.}
While our prompting framework is domain-agnostic and we have evaluated on a different dataset, broader validation is needed to confirm its wider applicability.

\section{Conclusion} 
We present a proactive approach to forecast toxicity detection in Open Source Software communities through early identification of conversational derailment on GitHub. We annotated and analyzed a dataset of 159 derailed toxic conversations and 207 non-toxic conversations. We developed a novel LLM-based prompt using Least-to-Most strategy to generate \textit{Summaries of Conversation Dynamics} and predict conversation derailment, achieving F1-scores of 0.901 with Qwen and 0.852 with Llama, significantly outperforming established baselines. External validation on an independent imbalanced dataset yielded F1-scores up to 0.797, demonstrating generalizability.

Our findings show that early derailment prediction is a feasible and effective strategy for moderating technical discussions. The explainable nature of our SCD-based approach enables transparent moderation decisions and flexible threshold-based interventions, allowing communities to shift from reactive to proactive moderation strategies. While our work has limitations regarding platform specificity and detection of sudden toxicity without warning signs, it provides the empirical foundation and practical tools necessary for implementing early warning systems in real-world OSS projects. Future work should extend to other platforms, incorporate behavioral signals, and evaluate live deployment and intervention outcomes to support healthier, more inclusive OSS communities. Another key direction can be generating human-written SCD in GitHub conversations and do instruct tune LLMs to mitigate the errors we observe at Section~\ref{sec:error} so that the LLMs can do better SCD generation.

% In this study, we aim to understand and predict conversational derailment and toxicity on GitHub. Our analysis reveals that users are more likely to initiate toxic conversations, while higher developer engagement correlates with lower toxicity levels. We also find that toxic conversations tend to be longer and escalate quickly after derailment. Generating conversation trajectory summaries using LLMs, we propose a proactive moderation approach, achieving F1-score of 0.70 with 0.76 precision in predicting conversational derailment.

% Future work should focus on enhancing our understanding of conversational derailment and toxicity on GitHub and similar platforms. Expanding the dataset to include conversations from additional OSS platforms like JIRA, Gitter and other discussion boards will help validate our findings across different environments and improve their generalizability. Developing and deploying real-time intervention tools that provide immediate feedback during conversations can prevent the escalation of toxic interactions. Understanding the most relevant social orientation tags in the OSS context can be another line of future work. 

\balance

\bibliographystyle{ACM-Reference-Format}
\bibliography{references}

\end{document}